\documentclass[12pt]{JHEP3}
\input epsf.tex
\epsfclipon

\usepackage{epsfig}
\usepackage{epstopdf}
\usepackage{epsf}
\input{epsf.sty}
\usepackage{graphicx,amsmath,amssymb}
\usepackage{epsfig,multicol}

\usepackage{bbm,bm,amsmath,amssymb}

\def\ba{\begin{eqnarray}}
\def\ea{\end{eqnarray}}

\newcommand{\roughly}[1]{\mathrel{\raise.3ex\hbox{$#1$\kern-0.85em
\lower1ex\hbox{$\sim$}}}}

\def\2pi{\left(2\pi\right)}

\def\beq{\begin{equation}}
\def\eeq{\end{equation}}
\def\bg{\begin{eqnarray}}
\def\nd{\end{eqnarray}}
\def\bea{\begin{eqnarray}}
\def\eea{\end{eqnarray}}

\def\D3{\overline{\mbox{D3}}}

\newcommand{\umax}{u_{\rm max}}

\newcommand{\xmax}{x_{\rm max}}
\newcommand{\ymax}{y_{\rm max}}

\def\blfootnote{\xdef\@thefnmark{}\@footnotetext}

\long\def\symbolfootnote[#1]#2{\begingroup%
\def\thefootnote{\fnsymbol{footnote}}\footnote[#1]{#2}\endgroup}

\newcommand{\be}{\begin{eqnarray}}
\newcommand{\ee}{\end{eqnarray}}
\newcommand{\ben}{\begin{eqnarray*}}
\newcommand{\een}{\end{eqnarray*}}

\newcommand{\bcent}{\begin{center}}
\newcommand{\ecent}{\end{center}}
\newcommand{\benum}{\begin{enumerate}}
\newcommand{\eenum}{\end{enumerate}}
\newcommand{\bdesc}{\begin{description}}
\newcommand{\edesc}{\end{description}}

\newcommand{\bitem}{\begin{itemize}}
\newcommand{\eitem}{\end{itemize}}

\newcommand{\bitemset}[1]{\begin{itemize}\addtolength{\itemsep}{#1}}
\newcommand{\biteml}{\begin{itemize}\addtolength{\itemsep}{+0.1cm}\large}
\newcommand{\bitemL}{\begin{itemize}\addtolength{\itemsep}{+0.1cm}\Large}
\newcommand{\bitemh}{\begin{itemize}\addtolength{\itemsep}{+0.1cm}\huge}
\newcommand{\bitemH}{\begin{itemize}\addtolength{\itemsep}{+0.1cm}\Huge}
\newcommand{\bitemn}{\begin{itemize}\addtolength{\itemsep}{+0.1cm}\normalsize}

\newcommand{\bquote}{\begin{quote}}
\newcommand{\equote}{\end{quote}}
\newcommand{\bhalfp}{\begin{minipage}{0.45\textwidth}}
\newcommand{\ehalfp}{\end{minipage}}
\newcommand{\bhead}{\begin{center}\bf \Large}
\newcommand{\ehead}{\end{center}\bigskip}
\newcommand{\non}{\nonumber\\}

 \newcommand{\calD}{{\cal D}}
 
 \newcommand{\calF}{{\cal F}}
 \newcommand{\calG}{{\cal G}}

\newcommand{\bmat}[1]{\left(\begin{array}{#1}}
\newcommand{\emat}{\end{array}\right)}

\usepackage{graphicx}

\title{Heavy Quarkonium Melting in Large $N$ Thermal QCD}

\author{Mohammed Mia, Keshav Dasgupta, Charles Gale, Sangyong Jeon\\
Department of Physics, McGill University,\\ 3600 University
Street, Montr{\'e}al QC, Canada H3A 2T8 }
\date{November 2009}

\abstract{Large $N$ QCD is mostly governed by planar diagrams and should show linear
confinement when these diagrams are suitably summed. The linear confinement 
of quarks in a class of these theories using gravity duals that capture the logarithmic runnings of the 
coupling constants in the IR and strongly coupled asymptotic
conformal behavior in the UV was studied in our previous work. We also extended the theories
to high temperatures and argued the possibilities of meltings and suppressions of heavy quarkonium states. 
In this paper we give a formal proof of melting using very generic choices of UV completions, and point out some 
subtleties associated with meltings in generic large $N$ theories.   
Our proof requires 
only the existence of well defined UV behaviors that are devoid of Landau poles and UV divergences of the 
Wilson loops, allowing degrees of freedom to increase monotonously with energy scale. We determine the 
melting temperatures of heavy quarkonium states, which could suggest the presence of deconfinement phase 
transitions in these theories.}

\maketitle

\begin{document}

\section{Introduction}

Large $N$ QCD in $3+1$ dimension
is a relatively simpler theory than its finite $N$ counterpart, owing in part to the $1/N$
suppressions of the non-planar diagrams, but its solution is still a challenge. For example its not clear how to 
add up all the planar diagrams and argue for linear confinement of fundamental quarks. Although in terms
of mesonic and glue-ball degrees of freedom the theory may look {\it free}, this is an oversimplification. Owing to 
the existence of a Hagedorn temperature and, at a more fundamental level, quark degrees of freedom, the system is 
still complex with no simple way of computing, for example, the current-current correlator or the master 
field \cite{thooft}. 

In principle, gauge-gravity duality provides a way to compute - or at least allow
some analytic control on -  some of these quantities. However
the problem is that to restrict everyting to the {\it supergravity} 
level, where we can have more analytic control, the gauge theory should be at strong 'tHooft coupling. This 
cannot happen for large $N$ QCD that is asymptotially free (although there could be a full string theory dual 
description). If one relaxes that condition and looks for strongly coupled asymptotic conformal behavior then
one can construct a supergravity dual for a class of these theories.   

In a previous paper \cite{jpsi}, which was a continuation of earlier work \cite{FEP},  
a new supergravity background that captures the logarithmic runnings of the coupling constants of a particular 
class of large $N$ QCD in the far IR and the strongly coupled asymptotic conformal behavior in the far UV was constructed. In the 
intermediate energy scales, our dual gravitational background captures the interpolating behavior of the beta 
function. 

In \cite{jpsi} it was argued that such a geometry would consist of three regions termed region 1, 2 and 
3 that would capture the IR, the intermediate scale, and the UV respectively. The supergravity solution was constructed
using fluxes sourced by $N$ number of D3 branes, $M$ number of D5 branes and anti-branes respectively while taking the
back reaction of $N_f$ number of seven branes and $N_f-3$ number of anti seven branes. The metric in all three regions 
can be written in the following form  \cite{jpsi}
\bg\label{bhmetko}
ds^2 = {1\over \sqrt{h}}
\Big[-g_1dt^2+dx^2+dy^2+dz^2\Big] +\sqrt{h}\Big[g_2^{-1}g_{rr}dr^2+ g_{mn}dx^m dx^n\Big]
\nd  
with $g_i$ being the black-hole factors and $h$ being the warp factor depending on all the internal 
coordinates ($r, \theta_i, \phi_i, \psi$).  To zeroth order in $g_sN_f$ and $g_s M$ we have our usual relations:
\bg\label{0gsnf}
&&h^{[0]} ~=~ {L^4\over r^4}, ~~~~g^{[0]} = 1-{r_h^4\over r^4}, ~~~~g^{[0]}_{rr} = 1, ~~~~
g^{[0]}_{mn}dx^m dx^n = ds^2_{T^{11}}\nonumber\\
&&g_1^{[0]}=g_2^{[0]}\equiv g= 1-\frac{r_h^4}{r^4}, ~~~~~~L^4\equiv g_s N \alpha'^2
\nd
where $T^{11}$ is the base of a six dimensional conifold and it has the topology of $S^3\times S^2$, where $S^3$ and $S^2$
are three-sphere and two-sphere. Here $g_s$ is the string coupling, $L$ is the AdS throat radius and $r_h$ is the black hole
horizon. The above metric is that of a ten dimensional geometry with coordinates $(t,x,y,z,r)$ describing a five dimensional
non compact space while internal coordinates ($\theta_1,\theta_2,\phi_1,\phi_2,\psi$) describe a five dimensional compact
space.

At higher orders in $g_sN_f,g_s M$, the warp factor, the black hole factors\footnote{This also means that the black hole
horizon is no longer the surface $r=r_h$ and the equation for the surface may be more involved.} and the internal metric get modified 
because of the back-reactions
from the seven-branes, three form fluxes and the localized sources we embed. We can write this as:
\bg\label{wfmet}
h ~ = ~ h^{[0]} ~+~ h^{[1]}, ~~~ g_{rr} ~=~ g_{rr}^{[0]} ~+~ g_{rr}^{[1]}, ~~~ g_{mn} ~=~ g^{[0]}_{mn} ~+~ 
g^{[1]}_{mn}, ~~~ g_i=g_i^{[0]}+ g_i^{[1]}
\nd
where the superscripts denote the order of $g_sN_f$ and $g_sM$. 

In regions 1 and 2, the warp factor $h$ has logrithmic dependence while in region 3, it can be expanded exclusively as a
power series in $1/r\equiv u$. In terms $u$ coordinate, we have the following Taylor series expansion for the warp factor $h$:
\bg \label{An}
{1\over \sqrt{h}}={\cal A}_n u^{n-2}
\nd  
where ${\cal A}_n$ are Taylor coefficients and in general can be functions of internal coordinates.
Region 3 was taken  to be large enough so that
the Nambu-Goto string would lie completely inside it. The computation of the potential at zero temperature 
revealed that at short distances the potential should be dominated by the Coulomb term\footnote{In deriving the
Coulomb potential, we have set AdS throat radius $L\equiv 1$ and string tension $T_0\equiv 1$ for convenience, and this 
is only a
redefinition of units. By restoring units, one obtains $V_{Q\bar{Q}} ~= ~ -\frac{0.236~\sqrt{g_s N}}{d} +{\cal O}(d)$, as expected
\cite{Mal-2}.}\cite{jpsi}:
\bg\label{sdpot}
V_{Q\bar{Q}} ~= ~ -{0.236\over d} ~ + ~ {\cal O}(d)
\nd
where the numerical value and the sign of the first term are fixed naturally by the dual background. It is 
interesting to note that this value is of the same order of magnitude as the one derived from QCD lattice 
simulations \cite{charmonium}. For large distances one expects the potential to be dominated by the linear term \cite{jpsi}:
\bg\label{lipo}
V_{Q\bar Q} ~ = ~ \left({{\cal A}_n {x}_{\rm max}^n \over \pi {x}_{\rm max}^2}\right) ~d
\nd 
where ${x}_{\rm max}$ is the maximum value for the depth of the U shaped string denoted by $u_{\rm max}$ and it is a
solution of the 
constraint equation (see section 2 and also \cite{jpsi} for details):
\bg \label{real-4}
{1\over 2}(m+1){\cal A}_{m + 3} {x}_{\rm max}^{m + 3} ~ = ~ 1
\nd 
The constraint equation is obtained by demanding  that the distance between the quarks and the potential energy of the 
quark pair is real. For AdS space, interquark distance $d$ and potential $V_{Q\bar{Q}}$ is {\it always} real for any value
of $u_{\rm max}$, and hence ${x}_{\rm max}$ do not exist, as there in {\it no} constraining equation.

At high temperatures and density, medium effects should {\it screen} the interaction between the heavy quark 
the anti-quark. The resulting effective potential between the quark anti-quark pairs separated by a distance $d$
at temperature ${T}$
can then be expressed 
succinctly in terms of the free energy $F(d, {T})$, which generically takes the following form:
\bg\label{freeenergy}
F(d, {T}) = \sigma d ~f_s(d, {T}) - {\alpha\over d} f_c(d, {T})
\nd
where $\sigma$ is the string tension, $\alpha$ is the gauge coupling and $f_c$ and $f_s$ are the screening 
functions\footnote{We expect the screening functions $f_s, f_c$ to equal identity when the temperature goes to 
zero, yielding the zero temperature Cornell potential.} 
(see for example \cite{karsch} and references therein). For the quark and the anti-quark pair kept at respective positions of 
$+{d\over 2}$ and $-{d\over 2}$, we expect the Wilson line $W\left(\pm {d\over 2}\right)$ to be related to the free 
energy through
\bg\label{wlfe}
{\rm exp}\left[-{F(d, {T})\over {T}}\right] ~ = ~ 
{\langle W^\dagger\left(+{d\over 2}\right) W\left(- {d\over 2}\right)\rangle \over 
\langle W^\dagger\left(+{d\over 2}\right)\rangle \langle W\left(-{d\over 2}\right)\rangle}
\nd
In terms of Wilson loop, the free energy Eq.\eqref{freeenergy} is now related to the renormalised Nambu-Goto 
action for the string on a background with a black-hole\footnote{There is a large body of literature on the 
subject where quark anti-quark potential has been computed using various different approaches like 
\cite{brambilla}. Although the potential Eq.\eqref{liponow} that we get matches well with other results, we have an 
additional constraint: Eq.\eqref{real-5}, from the background RG flow. This additional constraint, which cannot be 
seen from an AdS or an AdS with an IR cut-off dual, will have non-trivial consequences for melting that we will 
discuss in section 2.}. 
One may also note that the final theory is not three-dimensional, but four-dimensional and   
{\it compactified} on a circle in Euclideanised version.   

Using Eq.\eqref{wlfe} and the identification of the Wilson loop to the Nambu-Goto action,  
the free energy (or equivalently the potential) between the heavy
quark and the anti-quark at non-zero temperatures can be deduced. For large distances the potential is \cite{jpsi}
\bg\label{liponow}
V_{Q\bar Q} ~ = ~ {\sqrt{1-{{y}_{\rm max}^4\over {u}^4_{h}}}}\left({{\cal A}_n {y}_{\rm max}^n 
\over \pi {y}_{\rm max}^2}\right) ~d
\nd 
where ${y}_{\rm max}$ is the maximum value for the depth $u_{\rm max}$ of the U shaped string in the presence of a black
hole. It is  given by solving the constraint equation:
\bg\label{real-5}
{1\over 2} (m+ 1){\cal A}_{m+3} {y}_{\rm max}^{m+3}
+ {1\over j!}\prod_{k=0}^{j-1}\left(k-\frac{1}{2}\right)\left(\frac{{y}_{\rm max}^4}{u_h^4}\right)^j
\left[{\cal A}_{l}{y}_{\rm max}^{l}\left({l\over 2}+ 2j - 1\right)\right] ~= ~1\nonumber\\
\nd
which arises by demanding that inter quark separation and free energy
is always real \cite{jpsi}. Note that $u_h=1/r_h$ and for zero temperature, $u_h=\infty$. Thus at zero temperature, $u_h\rightarrow
\infty$ and  (\ref{real-5})
reduces to (\ref{real-4}).

At this point, it is important to define in a more precise manner what exactly is meant by melting. In this work, ``melting'' is meant to quantify the disappearance of the linear portion of the quark-antiquark potential, at a given length scale.  There is therefore a qualitative difference between this behavior and the actual dissolution of quarkonium bound states: The robustness of the $Q\bar{Q}$ pairs will actually depend on {\it where} their energy sits in the potential profile. Later studies could also involve a more precise characterization through a study of the temperature dependence of quarkonium spectral densities, for example. 

Going back to Eq.\eqref{liponow}, one might think that 
the melting temperature $u_c^{-1}$ is given by the condition
that $\ymax = u_h$ is a solution of Eq.(\ref{real-5}) so that the
coefficient of $d$ in Eq.(\ref{liponow}) vanishes.
Actually, this turned out {\it not} to be the case. 
As shown below, Eq.(\ref{real-5}) can never have a solution at
$\ymax=u_h$.
Hence, the deconfinement in our case does not mean a complete absence of 
the linear potential. Rather, it means that the linear potential ceases to
have an infinite range and as temperature increases, the range of the
linear
potential quickly becomes short provided that
Eq.\eqref{real-5} allows a real positive solution lying in
region 3. 
It is not clear this would always be the case, so Eq.\eqref{real-5}
requires a more detailed investigation. In the following section i.e sec. 2,  we
will carefully analyse Eq.\eqref{real-5} for generic choices of warp factors or more appropriately, generic UV completions
that have no Landau poles or UV divergences of the Wilson loops, and give a proof of quarkonium melting for this
class of theories. Section 3 contains a detailed numerical analysis that will allow us to find the 
melting points, and could also suggest the presence of a deconfinement phase transition in these 
theories. We conclude with a short discussion.

\section{Proof of the existence of a melting point}


Our cascading picture of renormalization group flow demands that 
in the region 3, the effective number of colors grows as $r$ grows.
The number of colors at any scale $u = 1/r$ in the region 3 is given by
$N_{\rm eff}(u) = N(1 + a_l u^l)$. 
For the analysis given here, it is simpler to define and use
\be
\calF(u) \equiv {u^2\over \sqrt{h}} = 
{\sqrt{N}\over L^2\sqrt{N_{\rm eff}}}
=
{1\over L^2\sqrt{1+ a_l u^l}}
\label{eq:F_of_u}
\ee
instead of $N_{\rm eff}(u)$. 
The coefficients ${\cal A}_n$ appearing in 
Eqs. (\ref{An}), (\ref{real-4}) and (\ref{real-5}) are related to $\calF(u)$ by
$\calF(u) = {\cal A}_n u^n$; and $h$ is the warp factor.
In terms of $\calF(u)$, the condition that $N_{\rm eff}(u)$ is 
a decreasing
function of $u=1/r$ becomes
\be
\calF'(u) > 0
\label{eq:fprime}
\ee
Combining Eqs.(\ref{eq:F_of_u}) and (\ref{eq:fprime})
yields the following condition 
\be
\calF(u) > {1\over L^2}
\ee
{}From now on, and as mentioned earlier, the value of $L$ is set to $1$ so that $\calF(u)> 1$. 

\subsection{Zero temperature}

Let $\umax$ be the maximum value of $u$ for the string between 
the quark and the anti-quark.
Then the relationship between $\umax$ and 
the distance between the quark and the anti-quark is given by \cite{jpsi}
\be
d(\umax) =
2\umax
\calF(\umax)
\int_{\epsilon_0}^1dv\,
{v^2 \sqrt{\calG_m \umax^m v^m}
\over (\calF(\umax v))^2}
\left[1-v^4\left(\calF(\umax)\over \calF(\umax v)\right)^2\right]^{-1/2}
\label{eq:dumax}
\ee
Two conditions must be met
before asserting that this expression represents the physical
distance between a quark and an anti-quark in vacuum. 
One obvious condition is that the integral is real.
This is guaranteed if for all   $0 \le v \le 1$:
\be
W(v|\umax) \equiv v^2\left( {\calF(\umax)\over \calF(\umax v)}\right) \le
1
\ee
Another condition is that the potential between the quark and the
anti-quark
must be long ranged. That is, $d(\umax)$ must range from 0 to $\infty$ as
$\umax$ varies from 0 to some finite value, say $\umax=\xmax$.
Since $\calF(u)> 1$, the only way to satisfy these conditions is via the (sufficiently fast) vanishing of the square-root
in Eq.(\ref{eq:dumax}) as $v\to 1$ at 
$\umax=\xmax$.

For most $\umax$, $1-W(v|\umax)^2$ vanishes only linearly as $v$ approaches
1.
In this case, $d(\umax)$ is finite since the singularity in the integrand
behaves like $1/\sqrt{1-v}$ and hence integrable.
To make $d(\umax)$ diverge at $\umax=\xmax$,
$1-W(v|\xmax)^2$ must vanish quadratically 
as $v$ approaches 1 so that the integrand is sufficiently singular,
$1/\sqrt{1-W(v|\xmax)^2} \sim 1/|1-v|$.
Therefore, the function $W(v|\xmax)$ must have a maximum at $v=1$.

To determine the value of $\xmax$, consider
\be
W'(v|\xmax)
& = &
2v \left( \calF(\xmax)\over \calF(\xmax v)\right)
\left(
1 - (\xmax v) {\calF'(\xmax v)\over 2\calF(\xmax v)}
\right)
\label{eq:Wprime}
\ee
For this to vanish at $v=1$, $\xmax$ must be 
the smallest positive solution of
\be
x\calF'(x) - 2\calF(x) = 0
\label{eq:zeroTcond}
\ee
With the definition $\calF(u) = {\cal A}_n u^n$, one can easily show that
this is equivalent to the condition (\ref{real-4}) which was originally
derived in \cite{jpsi}.
The allowed range of $\umax$ is then 
\be
0 \le \umax \le \xmax
\ee
and within this range, $d(\umax)$ varies from 0 to $\infty$.
How it varies will depend on the values of $\calG_m$ as well as
$\calF(u)$.

\subsection{Finite temperature}

At finite temperature, the relation between $\umax$ and the distance
between
the quark and the anti-quark is obtained by
replacing $\calF(u)$ with $\sqrt{1-u^4/u_h^4}\,\calF(u)$ in
Eq.(\ref{eq:dumax}):
\be
d_T(\umax) &=& 2\umax
\sqrt{1-\umax^4/u_h^4}\calF(\umax)
\int_{\epsilon_0}^1dv\,
{v^2 \sqrt{\calD_m \umax^m v^m}
\over (1-v^4\umax^4/u_h^4)(\calF(\umax v))^2}
\non & & {} \times
\left[
1-v^4{(1-\umax^4/u_h^4)\over(1-v^4\umax^4/u_h^4)}
\left(\calF(\umax)\over \calF(\umax v)\right)^2
\right]^{-1/2}
\label{eq:dumaxT}
\ee
The explicit factor of $\umax$ makes $d_T(\umax)$ vanish at $\umax=0$
as in the $T=0$ case.
As $\umax$ approaches $u_h$, the integral near $v=1$ behaves like
\be
d_T(\umax)
& \sim &
\int_0^1 dv
{\sqrt{1-\umax^4/u_h^4} \over \sqrt{(1-v)(1-v\umax/u_h)}}
\ee
%
%
which indicates that $d_T(\umax)$ goes to 0 as $\umax$ approaches $u_h$. 
Hence, at both $\umax=0$ and $\umax=u_h$, $d_T(\umax)$ vanishes.
Since $d_T(\umax)$ is positive in general,
there has to be a maximum between $\umax=0$ and $\umax=u_h$.
Whether the maximum value of $d_T(\umax)$ is infinite as in the $T=0$ case
depends on the temperature (equivalently, $u_h^{-1}$) as we now show.

The fact that the physical distance needs to be real
yields the following condition. For all $0 \le v \le 1$,
\be
W_T(v|\umax)\equiv
v^2\left(\calF(\umax)\over \calF(\umax v)\right)
\sqrt{1-\umax^4/u_h^4\over 1-\umax^4 v^4/u_h^4} \le 1
\ee
Taking the derivative gives
\be
W_T'(v|\umax)
& = &
{(1-\umax/u_h^4)^{1/2}\over(1-\umax^4 v^4/u_h^4)^{3/2}} 
{v\calF(\umax)\over\calF(\umax v)}
\non
& & {}\times
\left[
-(\umax v)(1-(\umax v/u_h)^4)\calF'(\umax v) + 2 \calF(\umax v)
\right]
\non
\label{eq:Wprime}
\ee
Similarly to the $T=0$ case, $d_T(\umax)$ can have an infinite range
if the derivative vanishes at
$v=1$ for a certain value of $\umax$, say $\umax = \ymax$.
This value of $\ymax$ is determined by
the smallest positive solution of the following equation
\be
y\calF'(y) - 2\calF(y) = (y/u_h)^4\,y\calF'(y) 
\label{eq:finiteTopt}
\ee
which then forces $W'_T(1|\ymax)$ to vanish.
Note that the left hand side is the same as the zero temperature
condition,
Eq.(\ref{eq:zeroTcond}). The right hand side is the temperature $(u_h)$
dependent part.
Using the facts that:
\bg\label{knownfact}
\prod_{k=0}^{j-1}\left(k-1/2\right)
= -(2j-3)!!/2^j,~~~ \sum_{j=1}^\infty x^j\, (2j-3)!!/2^j j!
= 1- \sqrt{1-x},
\nd
it can be readily
shown that Eq.(\ref{eq:finiteTopt}) is equivalent to Eq.(\ref{real-5}) as
long as $\umax < u_h$.
It is also clear that $y=u_h$ cannot be a solution of
Eq.(\ref{eq:finiteTopt}) because at $y=u_h$, the equation reduces to
$\calF(u_h) = 0$ which is inconsistent with the fact that $\calF(y)\ge 1$.

Recall that $\calF(y)\ge 1$ and $\calF'(y)\ge 0$ in the region 3 and we
assume that the equation $y\calF'(y)-2\calF(y)=0$ has a real positive
solution $\xmax$.
Hence as $y$ increases from 0 towards $\xmax$,
the left hand side of Eq.(\ref{eq:finiteTopt}) 
increases from $-2$ 
while the right hand side increases from 0.
The left hand side reaches 0 when $y = \xmax$
which is the point where the distance
$d(\umax)$ at $T=0$ becomes infinite.
At this point the right hand side of Eq.(\ref{eq:finiteTopt}) is
positive and has the value  
$(\xmax/u_h)^4\xmax\calF'(\xmax)$.
Hence the solution of Eq.(\ref{eq:finiteTopt}), if it exists,
must be larger than $\xmax$.

Consider first low enough temperatures so that $u_h \gg \xmax$.
For these low temperatures,
Eq.(\ref{eq:finiteTopt}) will have a solution, as the right hand side will be still small around $y =\xmax$.
This then implies that 
the linear potential at low temperature will have 
an infinite range if the zero temperature potential has an infinite range.

To show that the infinite range potential cannot be maintained at all 
temperatures, let $u_h = \xmax$.
When the left hand side vanishes at $y=\xmax$,
the right hand side is 
$\xmax\calF'(\xmax) = 2\calF(\xmax)$
which is positive and finite.
For $y > \xmax$, the left hand side ($y\calF'(y)-2\calF(y)$)
may become postive, but it is always smaller than $y\calF'(y)$ 
since $\calF(y)$ is always positive.
But for the same $y$, the right hand side ($(y/u_h)^4y\calF'(y)$)
is always positive and necessarily larger than
$y\calF'(y)$ since $(y/u_h) > 1$. 
Hence, Eq.(\ref{eq:finiteTopt}) cannot have a real and positive
solution when $u_h = \xmax$. 
Therefore between $u_h = \infty$ and $u_h = \xmax$, there must be a point
when Eq.(\ref{eq:finiteTopt}) cease to have a positive solution.

When Eq.(\ref{eq:finiteTopt}) has no solution, then the expression for
$d_T(\umax)$, (\ref{eq:dumaxT}) will not diverge for any $\umax$ 
within $(0, u_h)$.  Furthermore, since the expression vanishes at both
ends,
there must be a maximum $d_T(\umax)$ at a non-zero $\umax$. 
When the distance between the quark and the anti-quark is greater than
this
maximum distance, there can no longer be a string connecting the quark and
the anti-quark. 

\section{Numerical analyses of melting temperatures}

After discussing the most general choice for warp factors that give rise to $y_{\rm max}$ and 
consequently linear potential,  specific examples of geometries that may arise as solutions to Einstein's equation will be considered,  starting with the
following ansatz for the metric:
\bg \label{metric}
ds^2&=&-\frac{g}{\sqrt{h}}dt^2+\frac{1}{\sqrt{h}}(dx^2+dy^2+dz^2)+\frac{\sqrt{h}}{u^2}\left(\frac{H}{gu^2}du^2+ds^2_{{\cal M}_5}\right)\nonumber\\
&\equiv&-\frac{g}{\sqrt{h}}dt^2+\frac{1}{\sqrt{h}}(dx^2+dy^2+dz^2)+\frac{\sqrt{h}}{u^2}\widetilde{g}_{mn}dx^m dx^n
\nd
where $h\equiv h(u,\theta_i,\phi_i,\psi), H\equiv H(u,\theta_i,\phi_i,\psi)$, and 
$g\equiv 1-u^4/u_h^4$; ${\cal M}_5$ is the
compact five dimensional manifold parametrised by
coordinates $(\theta_i,\phi_i,\psi)$ and can be thought of as a
perturbation over $T^{1,1}$. Here $u=0$ is the boundary and $u=u_h$ is the horizon. 
As discussed in earlier work \cite{jpsi}, the above metric arises in region 3 of \cite{jpsi} when one considers the 
running of axio-dilaton $\tau$, $D7$ brane local action and fluxes due to 
anti five-branes on a
geometry that deviates from the IR OKS-BH geometry from the backreactions of the above sources.
The three-form fluxes sourced by ($p,q$) anti-branes are proportinal to $r^{-i}f(r)$ 
for some positive $i$ (see \cite{jpsi} for details about $f(r)$), where the function
 $f(r)\rightarrow 1$ as $r\rightarrow \infty$ and $f(r)\rightarrow 0$ as $r\rightarrow 0$. 
With the coordinate $u=1/r$, there is another function: $k(u) \equiv {\rm
 exp}(-u^{\cal A}), {\cal A}>0$, that also has somewhat similar behaviour as $f(u)$
and may allow us to have a better analytic 
control on the background. 
With such a choice of $k(u)$, the total three from flux is proportional to
 $u^A M(u)$ with 
\bg
M(u) \equiv M[1-k(u)] = M\left[1- {\rm exp}(-u^{\cal A})\right]
\nd
where $M$ is the number of bi-fundamental flavors.  
Thus three-form fluxes are decaying fast as 
$M u^A \left[1-{\rm exp}\left(-u^{\cal A}\right)\right]$
and, as shown in \cite{jpsi}, the seven-branes could be arranged such that the axio-dilaton 
$\tau$ behaves typically as 
$\tau \sim u^B$.
This means that
from the behavior of the internal Riemann tensor one may conclude that the internal metric 
$\widetilde{g}_{mn}$ behaves as
$\widetilde{g}_{mn}\sim u^C {\rm exp} (c_ou^{\cal C})$ 
where $A, C, {\cal A}$ and ${\cal C}$ are all positive and $c_o$ could be positive or negative 
depending on the precise background informations. 

{}From the above discussions it should be clear that taking the three-forms and world-volume gauge 
fluxes to be exponentially decaying in the IR
(but axio-dilaton to be suppressed only as $u^B$) should solve all the equations
of motion, giving the following behavior for the warp factor $h$ and the internal metric $H$ in 
\eqref{metric}\footnote{See also the interesting works of \cite{andreev} where exponential warp factors have 
been chosen.}:
\bg\label{warpy}
h~=~ L^4 u^4 {\rm exp}(-\alpha u^{\widetilde{\alpha}}), ~~~~~~ H~=~{\rm exp}(\beta u^{\widetilde{\beta}})
\nd    
where we are taking 
$\alpha, \widetilde{\alpha}, \beta, \widetilde{\beta}$ to be all positives with $\alpha, \beta$ to be 
functions of internal coordinates ($\theta_i,\phi_i,\psi$) and $L^4 = g_s N \alpha'^2$ to be the 
asymptotic AdS throat radius\footnote{Note that $\beta$ in \eqref{warpy} could be considered negative so that $H$ would 
be decaying to zero in the IR. However since region 3 doesn't extend to the IR we don't have to worry about the far IR 
behavior of Eq.\eqref{warpy}.}.  

Motivated by the above arguments, we will consider Nambu-Goto action of the string in the geometry with 
$(\widetilde{\alpha},
\widetilde{\beta})=(3,3)$ and $(\widetilde{\alpha},\widetilde{\beta})=(4,4)$ at temperatures $T^{(1)}$ and $T^{(2)}$
respectively in Eq.\eqref{warpy}. 
As in \cite{jpsi, FEP} we consider mappings
$X^\mu(\sigma, \tau)$, which are points in the internal space, to lie on the slice:
\bg\label{slice}
\theta_1~=~\theta_2~=~\pi, ~~~~~~ \phi_i~=~ 0, ~~~~~~ \psi~=~0
\nd
so that on this slice $\alpha, \beta$ are fixed and we set it to
$(\alpha,\beta)=(0.1, 0.05)$ for both choices  $(\widetilde{\alpha}, \widetilde{\beta})$. 
(Such a choice of slice will also help us to ignore the three-form contributions to the Wilson loop.)
With these fixed choices for the warp factors, we plot
the interquark separation $d$ as a function of $u_{\rm max}$  in figures 1 and 2
for various values of $T\equiv 1/u_h$.
\begin{figure}[htb]\label{dVSumax-1}
		\begin{center}
\includegraphics[height=9cm,angle=-90]{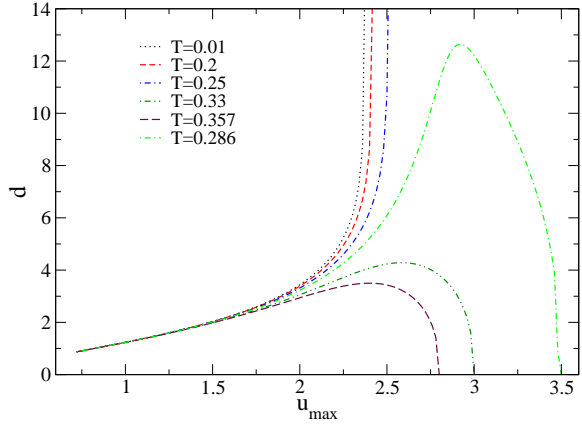}
		\caption{{Interquark distance as a function of $u_{\rm max}$ for various temperatures and warp factor with 
		$(\alpha,\widetilde{\alpha},\beta,\widetilde{\beta})=(0.1,3,0.05,3)$ in the warp factor equation.}}
		\end{center}
		\end{figure}
		
\begin{figure}[htb]\label{dVSumax-2}
		\begin{center}
\includegraphics[height=9cm,angle=-90]{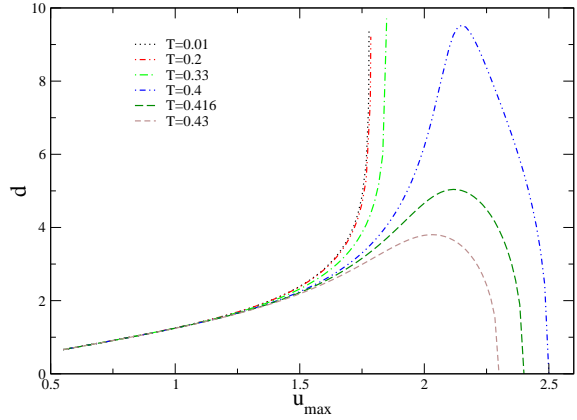}
		\caption{{Quark-antiquark distance as a function of $u_{\rm max}$ for various temperatures and warp factor with 
		$(\alpha,\widetilde{\alpha},\beta,\widetilde{\beta})=(0.1,4,0.05,4)$ in the warp factor equation.}}
		\end{center}
		\end{figure}
Note that for both choices of warp factors, for low enough temperatures, there exist $u_{\rm max}= y_{\rm max}$ where
$d\rightarrow \infty$. As the temperature is increased, $y_{\rm max}$ increases modestly. On the other hand from
figure \ref{dVSumax-1}, one sees that when $T>T_c^{(1)}\sim 0.28$ there 
exists a $d_{\rm max}$ which is finite. This means for
interquark distance $d > d_{\rm max}$, 
there is {\it no} string configuration with boundary condition $x(0)=\pm d/2$ implying that the
string attaching the quarks breaks
and we have two {\it free} partons for $d > d_{\rm max}$. Thus we can interpret $d_{\rm max}$
to be a ``screening length''. From figure 2 we observe similar behaviour but now $d_{\rm max}$ exists for
$T>T_c^{(2)}\sim 0.399$. 

\begin{figure}[htb]\label{dmaxvsT}
		\begin{center}
\includegraphics[height=9cm,angle=-90]{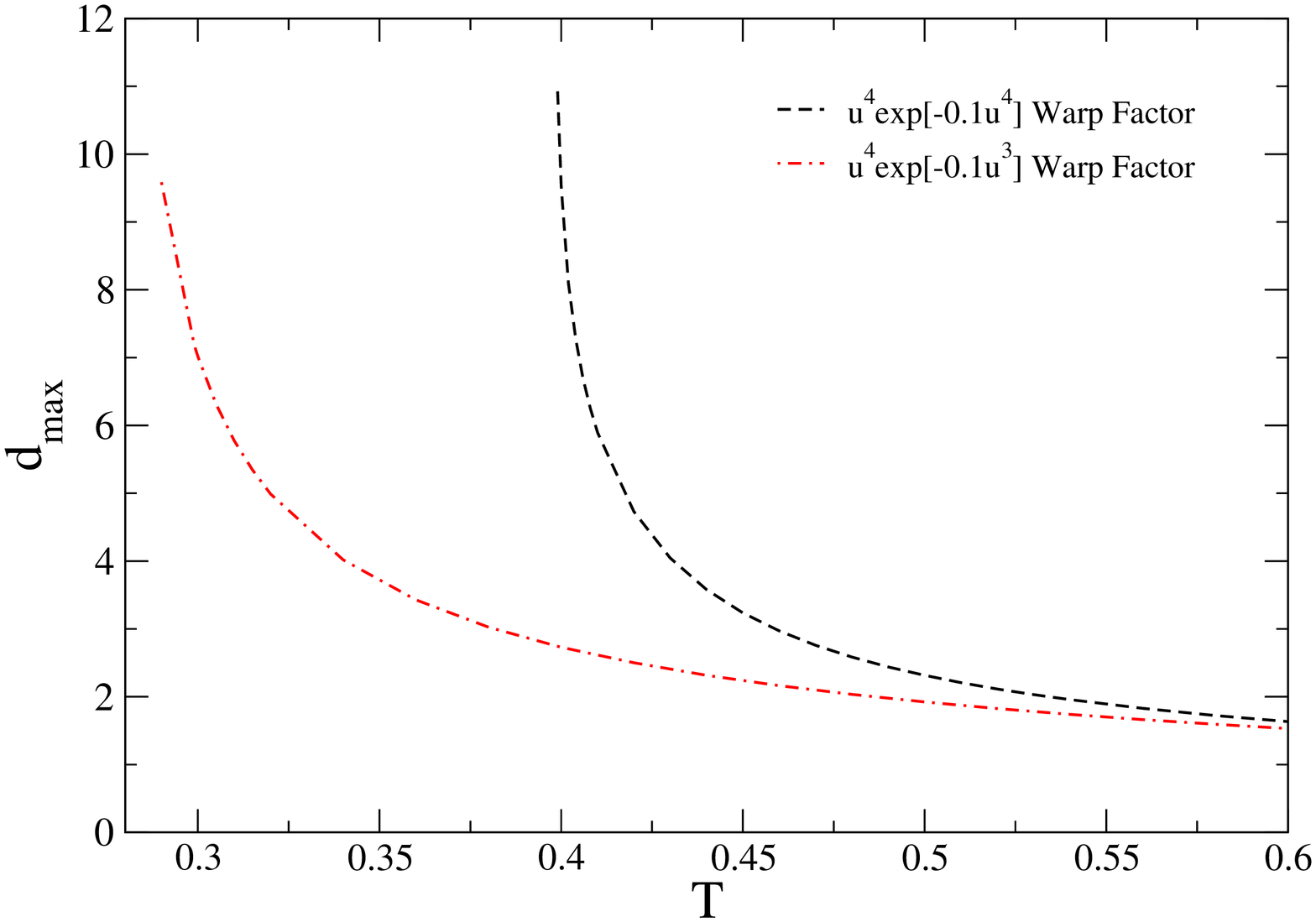}
		\caption{{Maximum interquark separation $d_{\rm max}$ as a function of $T=1/u_h$ for both cubic and 
quartic warp factors.}}
		\end{center}
		\end{figure}
In figure 3, $d_{\rm max}$ as a function of $T$ is plotted. We note that for a small change in the 
temperature near $T_c^{(1)}$ 
(or near $T_c^{(2)}$ equivalently) there is a sharp decrease in screening length 
$d_{\rm max}$, but for $T>> T_c^{(i)}$, $i = 1, 2$, 
 the screening length does not change much. In fact $d_{\rm max}$ behaves as
$C + {\rm exp}(-\gamma T)$ (where $C$ and $\gamma$ are constants) 
which in turn 
could be an indicative of a phase transition near $T_c^{(i)}$ for $i = 1,2$ i.e the two choices of warp factor. 
\begin{figure}[htb]\label{dmaxvsT}
		\begin{center}
\includegraphics[height=9cm,angle=-90]{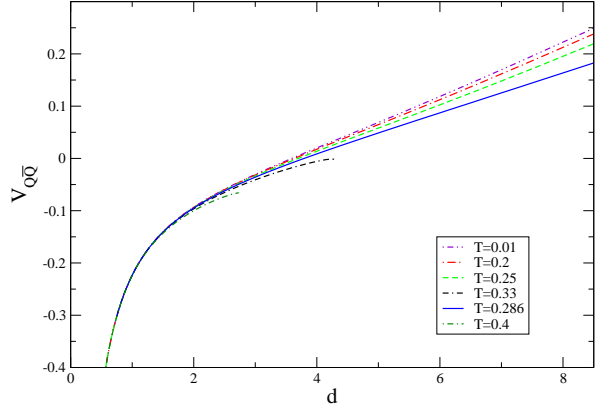}
		\caption{{Heavy quark potential $V_{Q\bar{Q}}$ as a function of quark separation $d$ with cubic warp 
factor, or equivalently,  
		$(\alpha,\widetilde{\alpha},\beta,\widetilde{\beta})=(0.1,3,0.05,3)$ in 
the warp factor equation for various temperatures. As mentioned in the text, one shouldn't consider the free energy
(or equivalently the potential) to stop abruptly in the plot. After the string connecting the quarks breaks, the curves  
should be extrapolated by the contributions to the potential energy from the two free strings {\it and} their 
world-volume fluctuations for all $T > T_c^{(i)}$.}}
		\end{center}
		\end{figure}
		
\begin{figure}[htb]\label{dmaxvsT}
		\begin{center}
\includegraphics[height=9cm,angle=-90]{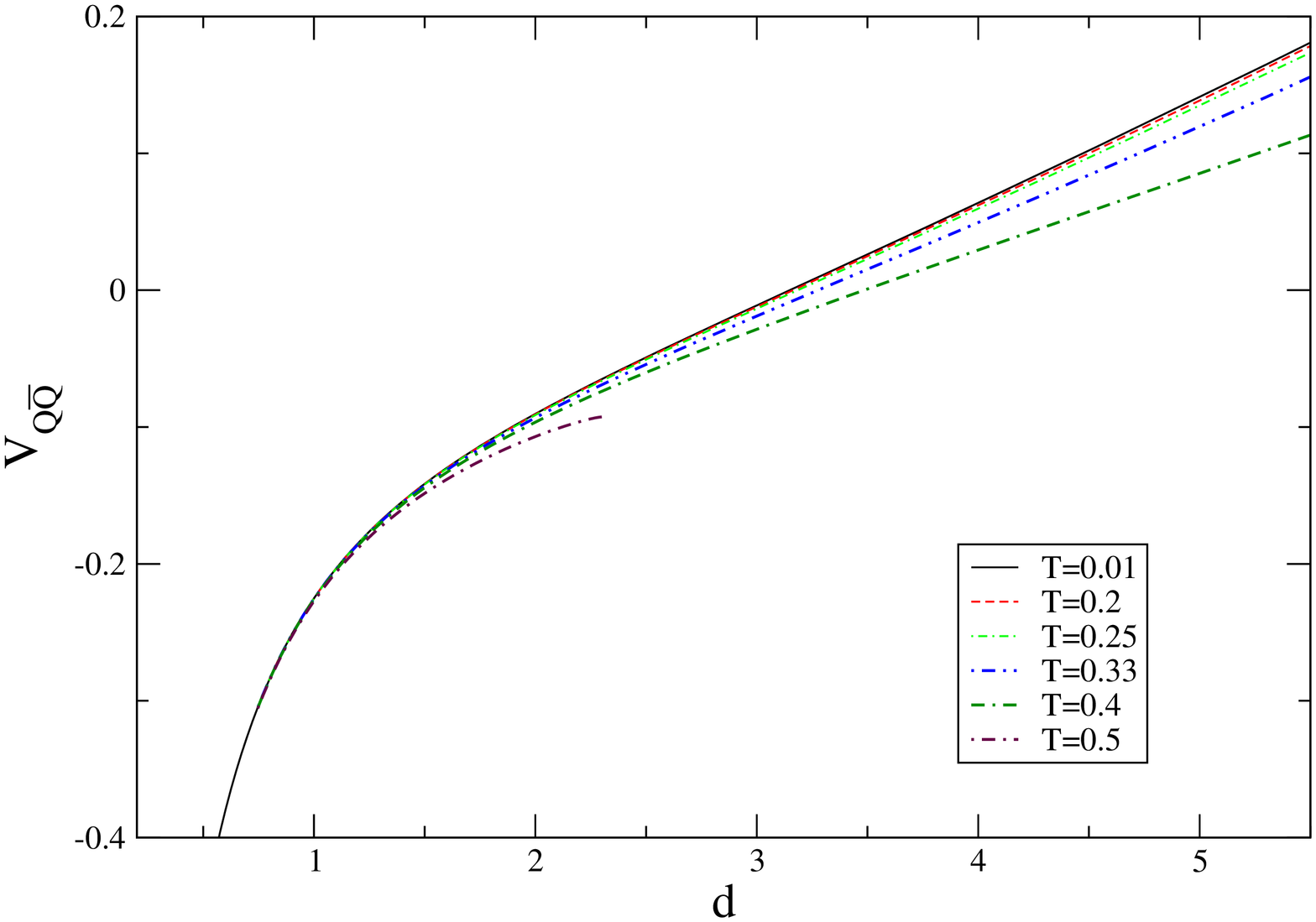}
		\caption{{Heavy quark potential $V_{Q\bar{Q}}$ as a function of quark separation $d$ with quartic warp 
factor, or equivalently,  
		$(\alpha,\widetilde{\alpha},\beta,\widetilde{\beta})=(0.1,4,0.05,4)$ in 
the warp factor equation for various temperatures. Here again, beyond $T > T_c$ the curves 
should be extrapolated by the contributions to the potential energy from the two free strings {\it and} their 
world-volume fluctuations.}}
		\end{center}
		\end{figure}
Finally we plot the potential $V_{Q\bar{Q}}$ as a function of $d$ in figures 4 and 5 for the two
choices of warp factor. For $T<T_c^{(1)}$ in figure 4 and $T<T_c^{(2)}$ in figure 5, we have potential energies
linearly increasing with an arbritrarily large increment of the 
interquark separations. Thus we have linear confinement of quarks for large
distances and small enough temperatures. For $T>T_c^{(i)}$, $i=1$ or $2$, there exists a $d_{\rm max}$ and for
all distances 
$d>d_{\rm max}$ there are no Nambu-Goto actions, $S_{\rm NG}$, for the
string attaching {\it both} the quarks. This means that we have free
quarks and $V_{Q\bar{Q}}$ is constant for $d>d_{\rm max}$. Of course looking at figures 4 and 5 one shouldn't 
conclude that the free energy {\it stops} abruptly. 
What happens for those two cases is that the string joining the quarks 
breaks, and then the free energy is given by the sum of the 
energies of the two strings (from the tips of the seven-branes to the 
black-hole horizon) and the total energies of the small fluctuations on the world-volume of the strings. The latter 
contributions are non-trivial to compute and we will not address these in any more detail here,
but energy conservations should tell us how to extrapolate the curves in figures 4 and 5, 
beyond the points where the string
breaks, for all $T > T_c$. Of course after sufficiently long time the two strings would dissipate their energies 
associated with their world-volume fluctuations and settle down to their lowest energy states.  

Observe that for a wide range of temperatures $0< T <T_c^{(i)}$, the
potential and thus the free energy hardly changes. But near a narrow range of temperatures $T^{(i)}_{c} - \epsilon
< T < T^{(i)}_{c} + \epsilon$ (where $\epsilon \sim 0.05$),
free energy changes significantly. For figure 5 the change is more abrupt than figure 4. This means as we go for 
bigger  
values of $\widetilde{\alpha}$, the change in free energy is sharper.

In figure 6, we plot the slope of the linear potential as a function of $T$. 
Again for a wide range of temperatures, there are no significant changes in the slope
but near $T_c^{(i)}$, the change is more dramatic: the slope decreases sharply, indicating again
the possibility of a phase transition near $T_c^{(i)}$.
As we noticed before, 
here too bigger exponent $\widetilde{\alpha}$ gives a sharper 
decline in the slope hinting that when $\widetilde{\alpha}>>1$, the transition would be more manifest. 

\begin{figure}[htb]\label{slope}
		\begin{center}
\includegraphics[height=9cm,angle=-90]{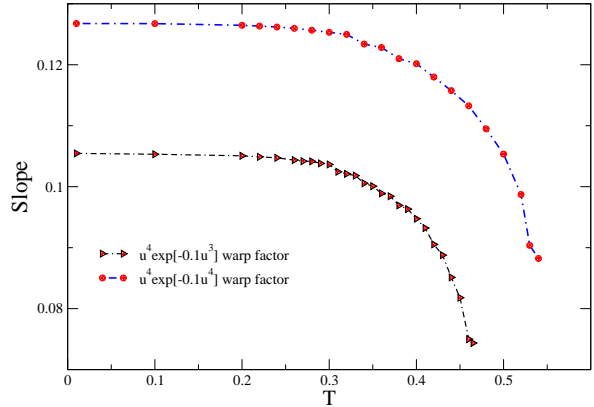}
		\caption{{Slope of linear potential as a function of T for both cubic and quartic warp factors. 
Note that in the figure the slopes have 
been computed as ${\Delta V\over \Delta d}$ with a range of $d$ from 1.6 to 1.7 in appropriate units.}}
		\end{center}
		\end{figure}
To conclude, the above numerical analyses suggest the presence of a deconfinement transition, where 
for a narrow range of temperatures $0.28 \le T_c  \le 0.39$ the free energy of 
$Q\bar{Q}$ pair shows a sharp decline. Interestingly, changing the powers of $u$ in the exponential 
changes the range of $T_c$ only by a small amount. So effectively $T_c$ lies in the range $0.2 \le T_c \le 0.4$. 
Putting back units, and defining the {\it boundary} temperature\footnote{See sec. (3.1) of \cite{FEP} for details.} 
${\cal T}$ as 
${\cal T} \equiv {g'(u_h)\over 4\pi\sqrt{h(u_h)}}$, 
our analyses reveal:
\bg
\frac{0.91}{L^2} ~\le ~{\cal T}_c ~\le ~ \frac{1.06}{L^2}
\nd
which is the range of the melting temperatures in these class of theories for heavy quarkonium states. Since the 
temperatures at both ends do not differ very much, this tells us that the melting temperature is inversely related to the 
asymptotic AdS radius in large $N$ thermal QCD.


\section{Discussions and conclusions}

In this work we attempted a small step towards the understanding of the melting temperatures of 
heavy quarkonium states in a class of large $N$ thermal QCD that have well defined UV behaviors without 
Landau Poles and UV divergences of the Wilson loops. Our analyses reveal two interconnecting results: any large 
$N$ gauge theory with $N_f$ fundamental quarks ($N_f \ll N$)
that is away from conformality with atmost one UV fixed point\footnote{This also means a good UV 
behavior with degrees of freedom increasing monotonously with energy scales.} 
should always have a mass gap and, consequently at a 
certain temperature $T_c$, the heaviest quarkonium states in such a theory should show melting. Such a conclusion seems 
to be true for generic non-supersymmetric theories, and it would be interesting to extend this to 
the supersymmetric cases. 

The melting temperatures\footnote{Note that their dependences on the 'tHooft coupling are consistent with what we 
would expect.}  
that we get for various UV completions seem not to differ too much from each other. 
This may mean that there 
is some underlying
universal behavior of heavy quarkonium states in large $N$ theories with good UV behaviors at strong 
'tHooft couplings. An interesting related study would be that of the behavior at weak 'tHooft couplings. There, there are also Hagedorn states that would come in because of the unsuppressed string modes. Details on this will be 
presented elsewhere.

\vskip1cm

\noindent{\bf Acknowledgements}

\noindent Its our pleasure to thank Sumit Das, Juan Maldacena, Guy Moore and Mark Van-Raamsdonk for helpful comments. 
This work was supported in part by the Natural Sciences and Engineering Research Council of Canada.

{}

\end{document}